%% file: doublePaper-1.tex
\documentclass[a4paper,12pt]{article}

\newcommand{\beqa}{\begin{eqnarray}}
\newcommand{\eeqa}{\end{eqnarray}}
\newcommand{\nn}{\nonumber}
\newcommand{\En}{E_{\nu_{l}i}}
\usepackage{epsfig}
\begin{document}
\begin{titlepage}
\title{Neutrino Masses or New Interactions}
\author{B.\ H.\ J.\ McKellar\thanks{e-mail b.mckellar@physics.unimelb.edu.au} 
\, and\, M.\ Garbutt\thanks{e-mail mag@physics.unimelb.edu.au} \\
University of Melbourne \\
Parkville, Victoria 3052, Australia\\
\\
G.\ J.\ Stephenson Jr.\ \thanks{e-mail: gjs@baryon.phys.unm.edu}\\
University of New Mexico \\
Albuquerque, New Mexico 87131 \\
\\
T.\ Goldman\thanks{e-mail goldman@t5.lanl.gov} \\
Los Alamos National Laboratory \\
Los Alamos, New Mexico 87545 }

\maketitle
\vspace*{-5.1in}
\flushright{LA-UR-01-3522}
\vspace{-.1in}
\flushright{UM-P-2001/017}
\vspace{-.4in}
\vspace*{4.9in}

\begin{abstract}

Recent proposals to study the mass of the ``electron" neutrino at a
   sensitivity of 0.3 eV can be used to place limits on the right handed
   and scalar charged currents at a level which improves on the present
   experimental limits. Indeed the neglect of the possibility of such
   interactions can lead to the inference of an incorrect value for the
   mass, as we illustrate.

\end{abstract}
\end{titlepage}

\section{Introduction}
Understanding the properties of the neutrino is one of the
foremost problems in modern particle physics, in both the experimental and
theoretical arenas.  Of prime importance is the determination of the
neutrino mass eigenstates: indications from neutrino oscillation
experiments are 
that at least one neutrino has a non-zero mass. The atmospheric
neutrino data measured at the Super-Kamiokande experiment favors
$\nu_\mu-\nu_\tau$ oscillation with a mass difference of $\delta
m_{atm}^2 \sim 3\times 10^{-3}$ \cite{Fukuda:1998mi}.  The simplest
interpretation of this is that one neutrino has a mass $m_3 \geq 0.05\
{\rm eV}$ and as such is the first hint of physics beyond
the Standard Model.  Neutrino oscillation experiments indicate the
finite value of the neutrino mass and the reveal how the mass
eigenstates are mixed, however they cannot determine the absolute
value of the neutrino masses, a question of equal 
significance.

The neutrino mass scale is accessible  through the
kinematics of weak interactions.  A program of Tritium
beta decay spectral measurements has been pursued to this end for a
number of years \cite{allofthem}.  Originally 
the focus of this program was to determine the Lorentz structure of 
Fermi's theory of weak interactions, that is Vector and  Axial-vector (V-A)
or some other 
combination of Scalar and Tensor currents\cite{number3}.  More recently the
focus has been solely on determining the mass of the electron
anti-neutrino through precision measurements of the end-point of the
electron energy spectrum \cite{Weinheimer:1999tn}~\cite{Lobashev:2001uu}.  The current upper bound on the neutrino mass set by
these experiments is $m_{\nu_e}\leq 2.5 {\rm eV}$
 while future experiments plan to achieve a sub-eV
sensitivity of $0.3\ {\rm eV}$, namely the KATRIN experiment to be
developed at Mainz \cite{Katrin} \cite{Lobashev:2001uu}.  The
interpretation of the upper bound to the neutrino mass must be made
with some care in light of the results from oscillation experiments,
since as previously noted, these experiments are
sensitive to the average mass of the mass eigenstates of the electron
neutrino state \cite{Vissani:2001ci} \cite{Farzan:2001cj}.  

Many extensions to the minimal Standard Model, as required by the
positive results of the oscillation experiments, introduce
interactions with Lorentz structures other than V-A which have a
coupling on a weaker scale than the Fermi coupling.  This possibility
must be included in the analysis of precision experiments such as the
next generation Tritium decay experiments; the implications of not
doing so have been examined previously in Refs.~(\cite{Stephenson:2000mw}) and (\cite{Stephenson:1998cx}) in the context
of the negative mass squared anomaly of the electron anti-neutrino. 
In this note we examine the consequences for the measured value of the
electron neutrino mass and the interpretation of this parameter in the
presence of non-standard currents. In particular we show that, in the 
presence of neutrino mixing, there may be interference effects which 
completely distort conclusions which may be drawn from the experiments 
concerning the nature of physics beyond the
Standard Model (SM).

\section{Formalism}
In this section we briefly outline a formalism used to describe
nuclear beta decay in the presence of interactions other than the 
standard model V-A, which is described in greater depth in 
Ref.~(\cite{Stephenson:2000mw}).   Low energy physics is best described
by a current-current 
type of interaction (also called four fermion contact interaction).
New physics at higher mass
scales will manifest itself as additional currents possibly with different
Lorentz symmetries from the dominant $V-A$ structure.  These
structures are $S,P,T,V$ and $A$, \emph{ie} Scalar, Pseudo-scalar, Tensor,
Vector and Axial Vector currents respectively, and can be recast into
their right and left
 handed components $S_R,S_L,T,R$ and $L$ with $L=(V-A)$,
$R=(V+A)$, $S_R=(S+P)$ and $S_L= (S-P)$.       

The most general effective interaction Hamiltonian for low energy,
semi-leptonic decays is given by,
\beqa
H_I = \sum_{\alpha,\beta = S_L,S_R,R,L,T}
G^{\alpha\beta}\sum_f\left(J_{h\alpha}^\dagger . J_{f\beta} + h.c.\right)\label{hamiltonian}
\eeqa
where $f=e,\ \mu,\  \tau,$ labels the weak eigenstate and in the Tritium case
$J_{f\lambda = e\lambda}=\bar{\psi}_e\Gamma_\lambda\psi_{\nu_e}$ represents the
leptonic current while $J_{h\alpha}$ represents the hadronic current.
The operators $\Gamma_\lambda$ are linear combinations of the five
bilinear covariants,
\beqa
\Gamma_{S_L}&=& (1-\gamma_5)\nonumber\\
\Gamma_{S_R} &=& (1+\gamma_5)\nonumber\\
\Gamma_{R} &=& \gamma^\mu(1+\gamma_5)\nonumber\\
\Gamma_{L} &=& \gamma^\mu(1-\gamma_5)\nonumber\\
\Gamma_{T} &=& [\gamma^\mu,\gamma^\nu]/2\ .
\eeqa
In the SM the only couplings present are $\beta = L$ and
$\alpha = L,R$.  These are expressed in more familiar notation as
\beqa
G^{RL} + G^{LL} = V_{ud}{\pi\alpha_W\over \sqrt{2} M_W^2}\ ,\label{smcoupling}
\eeqa   
where $V_{ud}$ is the element of the CKM matrix appropriate to nuclear beta
decay, $\alpha_W = {g_W^2\over 4\pi}$ is the fine structure
constant for the weak interaction of the SM related
to the coupling 
constant $g_W$, and $M_W$ is the mass of the $W^{\pm}$. 

The requirement that the Hamiltonian be Lorentz invariant causes most
off diagonal terms in the sum to vanish, with the exception being
$(S_R,S_L)$ and $(R,L)$.  
Note also that in the case of negatively charged lepton decay 
the subscripts of the operators $S_R$ and $S_L$ denote the chirality
of the neutrino field 
and do not correspond to the chirality of the negatively charged
lepton field.

Note also that in the case of $\beta^-$-decay
the subscripts of the operators $S_R$ and $S_L$ denote the chirality
of the neutrino field 
and do not correspond to the chirality of
electron field.    

The indications from neutrino oscillations are that the weak eigenstate
is not the same as the mass eigenstate.  We define the weak eigenstates
$\nu_f$ as linear combinations of the mass
eigenstates,
\beqa
\nu^f = \sum_i \cos\theta^f_i\ \nu_i
\eeqa
where the $cos\theta_i^f$ are the direction cosines in the coordinate
system spanned by the mass eigenstates.  

In general, should additional currents exist, the boson mediating the
current need not couple the eigenstates of the new interaction to
the same linear 
combination of mass eigenstates; hence we define
\beqa
\hat{\nu}^f = \sum_i \cos\theta^f_{iX}\ \nu_i
\eeqa
where the subscript `X' corresponds to the type of interaction and 
takes the values $S_R,\
S_L,\ R,$ or $T$. The angles `$\theta^f_{i}$' in the analysis of
oscillation experiments can 
be taken to lie in the first and fourth quadrant without loss of generality,
however relative to the new angles `$\theta^f_{iX}$' this is no
longer so.

The new Lorentz structures will produce interference terms in the beta
decay spectra proportional to the coupling constants
$G^{\alpha\beta}$.  It is these interference effects which are used to
produce the best limits on the strength of the new couplings
\cite{Adelberger:1999ud}. 
For a full description of the structure of the interference terms see
Ref.~(\cite{Stephenson:2000mw}).  Note also that the interference effects
due to the tensor (T) interaction with the left-chiral interaction are
the same as that of the scalar interactions and will not be discussed
further. 

The strength of interference effects are usually evaluated under the
assumption of that the weak eigenstate is the same as the mass
eigenstate, and quoted relative to the strength of the SM coupling
as,
\beqa 
\rho_X={\hat{g}^2M_W^2\over g^2M_X^2}
\eeqa
where $\hat{g}$ and $M_X$ are the coupling constant and mass of the
non-SM boson being exchanged and are related to the couplings of
Eq.~(\ref{hamiltonian}) by an analogous relationship to
Eq.~(\ref{smcoupling}).  The subscript $X$ is the type of interaction:
$S_R$, 
$S_L$ or $R$.   If the assumption that the weak and mass eigenstates
are the same is relaxed, the strength of new interactions must be
evaluated for each mass eigenstate on an individual basis.  The
constraints should now be taken to mean  $\rho_{\hat{X}} =\rho_X
\cos\theta_k\cos\theta_X$ .

Recent limits are given as \cite{Herczeg:1996qn} 
\beqa
\rho_R &\leq& 0.07\nn\\
\rho_{S_R}&\leq& 0.1\nn\\
\rho_{S_L}&\leq& 0.01\nn 
\eeqa 
where the equivalence of the mass and
weak eigenstates has been assumed.  The constraints on the left-chiral
scalar interaction are stronger than the others due to the fact that
this current would produce a charged lepton of the wrong chirality. 
Because of this strong constraint we do not consider this interaction
further.

\section{Spectra}
In this section we show how new Lorentz structures will be detected in
the various experimental spectra.
\subsection{Differential spectra and Kurie plots.}
The differential electron energy spectrum with scalar and right-handed
currents involves a sum over atomic final states, $i$, and is given by,
\beqa
\left({dN\over dE_\beta }({\cal E}^i_0)\right) & = & 
KF(E_\beta)q_\beta({\cal E}^i_0-E_\beta)\sum_{k}
\Theta({\cal E}^i_0-E_\beta-m_k)\nonumber\\
& \times&  E_\beta
({\cal E}^i_0-E_\beta )\sqrt{1-{m_k^2\over({\cal E}^i_0-E_\beta)^2}
}\nonumber\\
& \times & \biggl([\cos^2\theta_k + \cos^2\theta_{kR} {\rho_{R}}^2
+\biggl({G_V^2\over
 G_V^2+3 G_A^2} \biggr)\cos^2 \theta_{kS_R}\rho_{S_R}^2]\nonumber\\
&-&2m_em_k[\cos\theta_k\cos\theta_{kR}\rho_{R}]\nonumber\\
&-&m_kE_\beta \biggl({G_V^2\over
G_V^2+3G_A^2}\biggr)[\cos\theta_k\cos\theta_{kS_R}\rho_{S_R}]\biggr)\label{diff_spec}
\eeqa
The neutrino energy is defined by the difference between the end point
energy $({\cal E}_0^i)$ and the electron energy
$(E_\beta=\sqrt{q_\beta^2+m_\beta^2})$ as $E_\nu 
={\cal E}_0^i-E_\beta$.  The Fermi function is given by $F(E_\beta)$
and $m_k$ is the mass of the $k^{th}$ neutrino mass eigenstate.
Eq.~(\ref{diff_spec}) can be bought into a more convenient form by
noticing that near the end point the dependence on $E_\beta$ is very
weak and that the product $F(E_\beta)q_\beta $ is nearly constant.
Thus a new constant can be defined,
\beqa
K' = KF(E_\beta)q_\beta\ .
\eeqa 
Further define,
\beqa
\epsilon_{k R} = \rho_R{\cos\theta_{k R}\over \cos\theta_k}\ ,
\eeqa
and in the scalar case,
\beqa
\epsilon_{k S_R} = \rho_{S_R}{\cos\theta_{k S_R}\over \cos\theta_k}\ .
\eeqa
Now define,
\beqa
\epsilon_k =\epsilon_{k R}^2 + \epsilon_{k S_R}^2({G_V^2\over
G_V^2+3G_A^2}\biggr)
\eeqa
and finally 
\beqa
\phi_k = -2{m_k\over
(1+\epsilon_k)}\biggl[{m_e\over<E_\beta>}\epsilon_{kR}
+\epsilon_{kS_R}({G_V^2\over
G_V^2+3G_A^2}\biggr)\biggr]\ . \label{phi}
\eeqa
The ratio ${m_e\over E_\beta}$ varies by less than one part in a
thousand over the whole spectrum so can safely be included at its
average value in Eq.~(\ref{phi}).  The differential
spectrum is now
\beqa
\left({dN\over dE_\beta }({\cal E}^i_0)\right) & = &K'\sum_k 
(1+\epsilon_k)\cos^2\theta_k({\cal E}^i_0-E_\beta)^2(1+\epsilon_k)\left[1 +{\phi_k
\over({\cal E}^i_0-E_\beta) }\right]\nonumber\\
&\times& \sqrt{1-{m_k^2\over({\cal E}^i_0-E_\beta)^2}}\Theta({\cal
E}^i_0-E_\beta-m_k)\ .
\eeqa

\begin{figure}
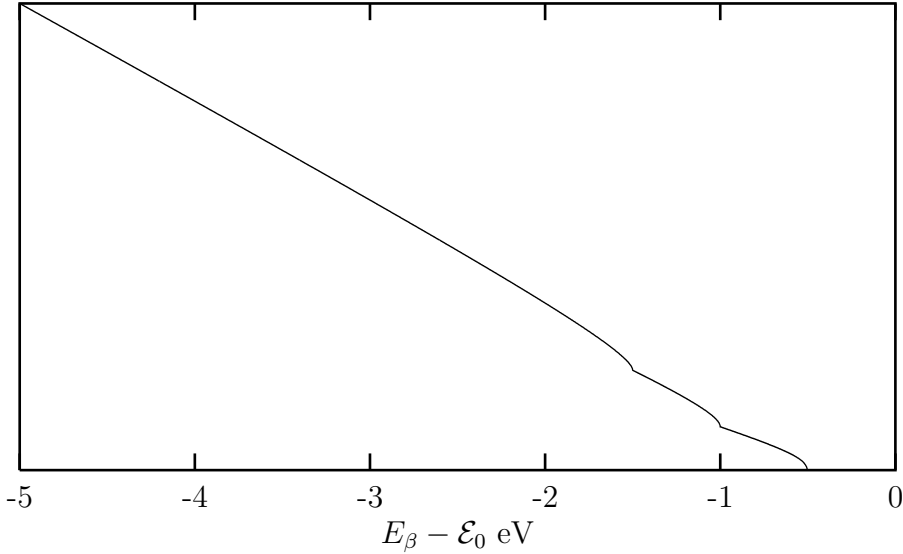

\include{kurie.ps}\label{kurie}
\caption{The last $5\ {\rm eV}$ of a Kurieplot for a fictitious mixing
scheme.  The three mass eignestates of $m_1=0.5$, $m_2=1.0$ and
$m_3=1.5\ {\rm eV}$ are maximally mixed.}
\end{figure}

The effects of a finite neutrino mass on the differential spectrum is to cause
a distortion near the end point, as the spectrum falls off as
$E_\nu q_\nu$ rather than $E_\nu^2$, and to shift the end point by
$m_\nu\ ({\cal E}_0\to {\cal E}_0-m_\nu)$. Neutrino mixing has a
similar result, see Fig.~\ref{kurie}, producing several kinks in the spectrum at the point
where the electron energy is such that decays to the $k^{th}$ mass
eigenstate are kinematically disallowed
\cite{McKellar:1980cn}~\cite{Shrock:1980vy}.  In what follows, 
we refer to $\frac{dN}{dE_{\beta}}({\cal E}^i_0)$ as
$\frac{dN_i}{dE_{\beta}}$.

\subsection{Integral spectra}
In the past many of the Tritium decay experiments performed were differential
measurements.  However, more recently and in the future, experiments
measure an integral spectrum.  That is, for each of the `$i$'
atomic/molecular final
states they count the number of electrons
above some cut off energy $E_\beta^C$,
\beqa
N_i(E_\beta^C)=\int^\infty_{E_\beta^C}{{\rm d}N_i\over {\rm
d}E_\beta}{\rm d}E_\beta\ .
\eeqa
In the Mainz experiment results for a number of different values of
$E_\beta^C$ are combined in a weighted average; the data is
then fitted as a function of  $E_l < E_\beta^C$, a lowest cutoff energy.
The task of weighting the average takes into account the
systematics of the spectrometer \cite{Weinheimer:1999tn}.  While we do
not attempt to reproduce the details of this calculation we can gain
insight from a crude model of this process, in which we integrate the integral
spectrum from $E_l$ to infinity,
\beqa
G_i(E_l)&=&\int^\infty_{E_l} N_i(E_\beta^C)dE_\beta^C\nn\\
&=&K'\sum_k (1+\epsilon_k)\cos^2\theta_k\Biggl({1\over
12}\biggl[\En(\En^2-m_k^2)^{3 \over 2} - {3 \over 2}m_k^2\En
(\En^2-m_k^2)^{1\over 2} \nn\\
& + &  {3 \over 2}m_k^4 {\rm cosh}^{-1}(\En/m_k)\biggr]  
+ {\phi_k\over 2}\biggl[{1\over 3} (\En^2-m_k^2)^{3 \over 2} \nn\\
 &-&m_k^2\En {\rm cosh}^{-1}(\En/m_k)-m_k^2 (\En^2-m_k^2)^{1\over
2}\biggr]\Biggr) 
\eeqa 
where in the last expression we have changed variables to $\En = {\cal
E}^i_0 -E_l$, representing the neutrino energy in the $i^{th}$
channel.  In order to gain some understanding of the behavior of the
equation we restrict our investigations to neutrino energies well below
the end point, $\En \gg m_k$, with this approximation the double
integral becomes,
\beqa
G_i(E_l) \approx {K'\over 12}\sum_k(1+\epsilon_k)
\En^4\biggl[1-3{m_k^2\over \En^2} + 2{\phi_k\over \En}\biggr]\
.\label{double_sim}
\eeqa
The approximation is accurate to about one part in a thousand for
cutoff energies $\En>35\ {\rm eV}$, and deteriorates
rapidly for energies $\En < 10 {\rm eV}$.
To simplify the discussion we examine the case of a decay to one
final state of the atomic/molecular system by setting $i=0$.  Further we
restrict the discussion to only one mass eigenstate $k=1$.  With these
simplifications the function to be fitted to the data will be
\beqa
G_0(E_l) \approx {{\cal A}\over 12}\biggl[\En^4 -3\En^2 m_1^2
+ 2\En^3 {\phi_1} + {\cal B}\En\biggr]\ .
\eeqa
The coefficients of the mass and interference terms will not, due to
the approximation made in Eq.~(\ref{double_sim}), be
exactly $2$ and $-3$; they will however be fixed and are not free to
be fitted.  The overall amplitude ${\cal A}=
(1+\epsilon_1)\cos^2\theta_1$ will be determined from the data, as
will the end point ${\cal E}^0_0$. The background ${\cal B}$ can be
independently determined from measurements far from the end point.  

From a set of a measured values of $G_0(E_l)$ for different $E_{l}$,
and for fixed ${\cal A}$, ${\cal B}$ and ${\cal E}^0_0$ the contours
of equal likelihood will trace approximate ellipses in the $(m_1^2,
\phi_1)$ plane.  The major axes of the ellipses will have a positive
slope as, at fixed $E_{l}$ the doubly integrated spectrum is sensitive
to $1.5 m_{1}^{2} - \phi_{1}E_{l}$, rather than the orthogonal
combination.  As different values of the parameters ${\cal A}$, ${\cal
B}$ and ${\cal E}^0_0$ are tried, the center of the ellipse, as
projected on to the $(m_1^2, \phi_1)$ plane moves over the plane.  The
curves of equal likelihood are thus going to be quite complex if the
parameters ${\cal A}$, ${\cal B}$ and ${\cal E}^0_0$ are optimized at
fixed $m_{1}^{2}, \phi_{1}$, and the equal likelihood contours plotted
in these remaining variables.  

In fitting the data it is important to be sure that all of the 
parameters, including  ${\cal A}$, ${\cal B}$ and ${\cal E}^0_0$  are
physically reasonable, and then  to find a minimum $\chi^2$ at a point
on the $(m_1^2, \phi_1)$ plane consistent with the independent limits
on $\epsilon_k$.  Discussions of the experimental results of
future
experiments should include a discussion of the values of all of the
fitted parameters.

\begin{figure}
\centering
\epsfig{file=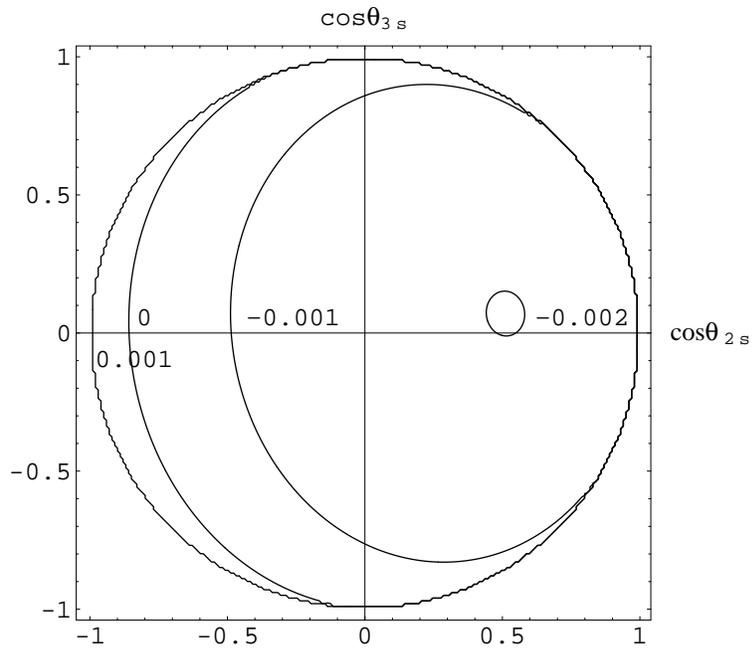, width=10cm}
\caption{Constant value contours of $\Delta G$ for the right-chiral
scalar interaction as compared to the spectrum produced assuming no
additional interactions and the Large Mixing Angle solution of the
solar neutrino 
data.  In this plot the third mass eigenstate has a mass of $m_3=1\
{\rm eV}$. The numbers immediately to the right of each contour give
the value of $\Delta G$.}
\label{scalar}
\end{figure}

\begin{figure}
\centering
\epsfig{file=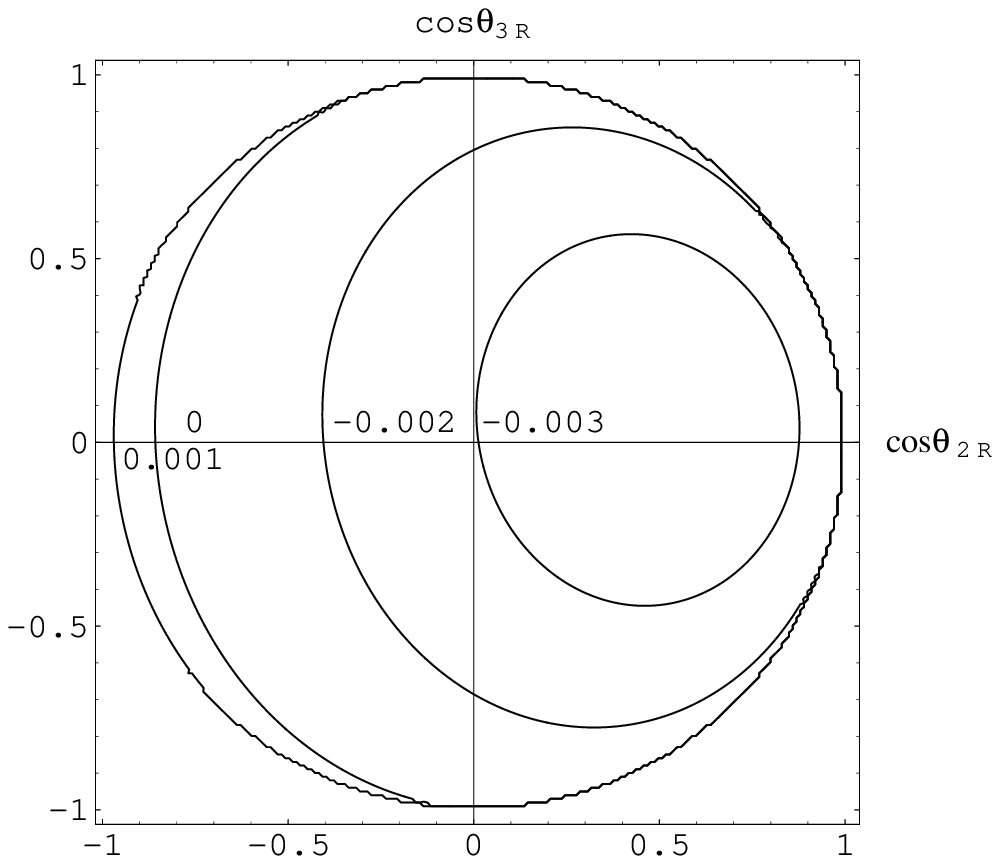, width=10cm}
\caption{Constant value contours of $\Delta G$ for the right-handed
vector interaction as compared to the spectrum produced assuming no
additional interactions and the Large Mixing Angle solution of the
solar neutrino 
data.  In this plot the third mass eigenstate has a mass of $m_3=1\
{\rm eV}$. The numbers immediately to the right of each contour give
the value of $\Delta G$.}
\label{right}
\end{figure}

We now generalize to the case where all mass eigenstates
participate in the decay.
The KATRIN experiment should be sensitive to a mass of
$0.3\ {\rm eV}$, or about a $0.02\%$ shift in $G_0(E_l)$ at an energy
of $\En\approx40\ {\rm 
eV}$.  For the various solutions to the solar and atmospheric neutrino
problem this threshold may be reached depending on the mass scale
\cite{Farzan:2001cj}. 
The influence of additional interactions, however, is uncertain.  We
investigate this by assuming the solutions to
the oscillation data fix the values of $\cos\theta_k$
and also set the mass differences between the mass eigenstates.  The
value of $G_0(\En)$ will then be determined by the coupling strength and
mixing of the additional interactions.
To investigate the magnitude of this influence
the absolute value of the  relative difference between the spectra
with and without ($G_0^{X}$ and $G_0'$  respectively)
right-chiral scalar and
right-handed vector couplings over the space of the non-standard direction
cosines have been plotted.  That is, contours of
\beqa
\Delta G  = {G_0^X-G_0'\over G_0' }
\eeqa  
in the $(\cos\theta_{2 X},\ \cos\theta_{3 X})$ plane are shown in Figs.~\ref{scalar} and \ref{right}.  In this example the standard direction
cosines are defined by the Large Mixing Angle solution to the solar
neutrino problem, as are the mass differences.  

The mass scale is chosen so as not to
conflict with cosmological bounds \cite{Pas:2001nd};  in this context it would be
inappropriate to use bounds from the Tritium decay experiments.
In producing these plots unitarity
for both the standard and non-standard mixing cosines has been
assumed. The amplitude for the spectra will be
experimentally determined, and as such, will be independent of
assumptions made about 
the additional interactions; it is defined as ${\cal A} =
K'\sum_k(1+\epsilon_k^2)\cos\theta_k^2$.   

We note that for some values of $\cos{\theta_{k X}}$ there is no
difference between the standard and non-standard spectra as a result
of destructive interference between different mass eigenstates.  
This is an important point to consider in discussions about
constraints on mass
scales of new physics derived from the non-observation of interference
effects in spectra or correlation parameters.
The magnitude of the relative difference between spectra for some
values $\cos{\theta_{k X}}$ 
is larger than the $0.02\%$ detection threshold.  An assumption of a
particular standard mixing scheme in this situation would result in an
incorrect fit for the value of $m_{\nu_e}$.  This reinforces the need
for the possibility of weaker interactions in beta-decay to be taken into
account when analysing future precision results.    

\section{Conclusion}
In this note we have tried to highlight the important role that
interactions other than the dominant $V-A$ interaction may play
in precision measurements of the Tritium beta decay spectrum.  If
these are not
correctly accounted for, fits to the neutrino mass will be misleading,
and a door to new physics may be closed.  The model
for the integral spectra we have  presented gives some insight into
the likely effects of new interactions.  Discussions of future experimental
results at the expected level of sensitivity should include a full
account of fits with the interference 
terms incorporated into the analysis.     
\section*{Acknowledgments}
This research is supported in part by the Department of Energy under
contract W-7405-ENG-36, in part by the Australian Research Council and in
part by the National Science Foundation.

\end{document}

%% file: kurie.ps.tex
\begingroup%
  \makeatletter%
  \newcommand{\GNUPLOTspecial}{%
    \@sanitize\catcode`\%=14\relax\special}%
  \setlength{\unitlength}{0.1bp}%
{\GNUPLOTspecial{!
/gnudict 256 dict def
gnudict begin
/Color false def
/Solid false def
/gnulinewidth 5.000 def
/userlinewidth gnulinewidth def
/vshift -33 def
/dl {10 mul} def
/hpt_ 31.5 def
/vpt_ 31.5 def
/hpt hpt_ def
/vpt vpt_ def
/M {moveto} bind def
/L {lineto} bind def
/R {rmoveto} bind def
/V {rlineto} bind def
/vpt2 vpt 2 mul def
/hpt2 hpt 2 mul def
/Lshow { currentpoint stroke M
  0 vshift R show } def
/Rshow { currentpoint stroke M
  dup stringwidth pop neg vshift R show } def
/Cshow { currentpoint stroke M
  dup stringwidth pop -2 div vshift R show } def
/UP { dup vpt_ mul /vpt exch def hpt_ mul /hpt exch def
  /hpt2 hpt 2 mul def /vpt2 vpt 2 mul def } def
/DL { Color {setrgbcolor Solid {pop []} if 0 setdash }
 {pop pop pop Solid {pop []} if 0 setdash} ifelse } def
/BL { stroke userlinewidth 2 mul setlinewidth } def
/AL { stroke userlinewidth 2 div setlinewidth } def
/UL { dup gnulinewidth mul /userlinewidth exch def
      10 mul /udl exch def } def
/PL { stroke userlinewidth setlinewidth } def
/LTb { BL [] 0 0 0 DL } def
/LTa { AL [1 udl mul 2 udl mul] 0 setdash 0 0 0 setrgbcolor } def
/LT0 { PL [] 1 0 0 DL } def
/LT1 { PL [4 dl 2 dl] 0 1 0 DL } def
/LT2 { PL [2 dl 3 dl] 0 0 1 DL } def
/LT3 { PL [1 dl 1.5 dl] 1 0 1 DL } def
/LT4 { PL [5 dl 2 dl 1 dl 2 dl] 0 1 1 DL } def
/LT5 { PL [4 dl 3 dl 1 dl 3 dl] 1 1 0 DL } def
/LT6 { PL [2 dl 2 dl 2 dl 4 dl] 0 0 0 DL } def
/LT7 { PL [2 dl 2 dl 2 dl 2 dl 2 dl 4 dl] 1 0.3 0 DL } def
/LT8 { PL [2 dl 2 dl 2 dl 2 dl 2 dl 2 dl 2 dl 4 dl] 0.5 0.5 0.5 DL } def
/Pnt { stroke [] 0 setdash
   gsave 1 setlinecap M 0 0 V stroke grestore } def
/Dia { stroke [] 0 setdash 2 copy vpt add M
  hpt neg vpt neg V hpt vpt neg V
  hpt vpt V hpt neg vpt V closepath stroke
  Pnt } def
/Pls { stroke [] 0 setdash vpt sub M 0 vpt2 V
  currentpoint stroke M
  hpt neg vpt neg R hpt2 0 V stroke
  } def
/Box { stroke [] 0 setdash 2 copy exch hpt sub exch vpt add M
  0 vpt2 neg V hpt2 0 V 0 vpt2 V
  hpt2 neg 0 V closepath stroke
  Pnt } def
/Crs { stroke [] 0 setdash exch hpt sub exch vpt add M
  hpt2 vpt2 neg V currentpoint stroke M
  hpt2 neg 0 R hpt2 vpt2 V stroke } def
/TriU { stroke [] 0 setdash 2 copy vpt 1.12 mul add M
  hpt neg vpt -1.62 mul V
  hpt 2 mul 0 V
  hpt neg vpt 1.62 mul V closepath stroke
  Pnt  } def
/Star { 2 copy Pls Crs } def
/BoxF { stroke [] 0 setdash exch hpt sub exch vpt add M
  0 vpt2 neg V  hpt2 0 V  0 vpt2 V
  hpt2 neg 0 V  closepath fill } def
/TriUF { stroke [] 0 setdash vpt 1.12 mul add M
  hpt neg vpt -1.62 mul V
  hpt 2 mul 0 V
  hpt neg vpt 1.62 mul V closepath fill } def
/TriD { stroke [] 0 setdash 2 copy vpt 1.12 mul sub M
  hpt neg vpt 1.62 mul V
  hpt 2 mul 0 V
  hpt neg vpt -1.62 mul V closepath stroke
  Pnt  } def
/TriDF { stroke [] 0 setdash vpt 1.12 mul sub M
  hpt neg vpt 1.62 mul V
  hpt 2 mul 0 V
  hpt neg vpt -1.62 mul V closepath fill} def
/DiaF { stroke [] 0 setdash vpt add M
  hpt neg vpt neg V hpt vpt neg V
  hpt vpt V hpt neg vpt V closepath fill } def
/Pent { stroke [] 0 setdash 2 copy gsave
  translate 0 hpt M 4 {72 rotate 0 hpt L} repeat
  closepath stroke grestore Pnt } def
/PentF { stroke [] 0 setdash gsave
  translate 0 hpt M 4 {72 rotate 0 hpt L} repeat
  closepath fill grestore } def
/Circle { stroke [] 0 setdash 2 copy
  hpt 0 360 arc stroke Pnt } def
/CircleF { stroke [] 0 setdash hpt 0 360 arc fill } def
/C0 { BL [] 0 setdash 2 copy moveto vpt 90 450  arc } bind def
/C1 { BL [] 0 setdash 2 copy        moveto
       2 copy  vpt 0 90 arc closepath fill
               vpt 0 360 arc closepath } bind def
/C2 { BL [] 0 setdash 2 copy moveto
       2 copy  vpt 90 180 arc closepath fill
               vpt 0 360 arc closepath } bind def
/C3 { BL [] 0 setdash 2 copy moveto
       2 copy  vpt 0 180 arc closepath fill
               vpt 0 360 arc closepath } bind def
/C4 { BL [] 0 setdash 2 copy moveto
       2 copy  vpt 180 270 arc closepath fill
               vpt 0 360 arc closepath } bind def
/C5 { BL [] 0 setdash 2 copy moveto
       2 copy  vpt 0 90 arc
       2 copy moveto
       2 copy  vpt 180 270 arc closepath fill
               vpt 0 360 arc } bind def
/C6 { BL [] 0 setdash 2 copy moveto
      2 copy  vpt 90 270 arc closepath fill
              vpt 0 360 arc closepath } bind def
/C7 { BL [] 0 setdash 2 copy moveto
      2 copy  vpt 0 270 arc closepath fill
              vpt 0 360 arc closepath } bind def
/C8 { BL [] 0 setdash 2 copy moveto
      2 copy vpt 270 360 arc closepath fill
              vpt 0 360 arc closepath } bind def
/C9 { BL [] 0 setdash 2 copy moveto
      2 copy  vpt 270 450 arc closepath fill
              vpt 0 360 arc closepath } bind def
/C10 { BL [] 0 setdash 2 copy 2 copy moveto vpt 270 360 arc closepath fill
       2 copy moveto
       2 copy vpt 90 180 arc closepath fill
               vpt 0 360 arc closepath } bind def
/C11 { BL [] 0 setdash 2 copy moveto
       2 copy  vpt 0 180 arc closepath fill
       2 copy moveto
       2 copy  vpt 270 360 arc closepath fill
               vpt 0 360 arc closepath } bind def
/C12 { BL [] 0 setdash 2 copy moveto
       2 copy  vpt 180 360 arc closepath fill
               vpt 0 360 arc closepath } bind def
/C13 { BL [] 0 setdash  2 copy moveto
       2 copy  vpt 0 90 arc closepath fill
       2 copy moveto
       2 copy  vpt 180 360 arc closepath fill
               vpt 0 360 arc closepath } bind def
/C14 { BL [] 0 setdash 2 copy moveto
       2 copy  vpt 90 360 arc closepath fill
               vpt 0 360 arc } bind def
/C15 { BL [] 0 setdash 2 copy vpt 0 360 arc closepath fill
               vpt 0 360 arc closepath } bind def
/Rec   { newpath 4 2 roll moveto 1 index 0 rlineto 0 exch rlineto
       neg 0 rlineto closepath } bind def
/Square { dup Rec } bind def
/Bsquare { vpt sub exch vpt sub exch vpt2 Square } bind def
/S0 { BL [] 0 setdash 2 copy moveto 0 vpt rlineto BL Bsquare } bind def
/S1 { BL [] 0 setdash 2 copy vpt Square fill Bsquare } bind def
/S2 { BL [] 0 setdash 2 copy exch vpt sub exch vpt Square fill Bsquare } bind def
/S3 { BL [] 0 setdash 2 copy exch vpt sub exch vpt2 vpt Rec fill Bsquare } bind def
/S4 { BL [] 0 setdash 2 copy exch vpt sub exch vpt sub vpt Square fill Bsquare } bind def
/S5 { BL [] 0 setdash 2 copy 2 copy vpt Square fill
       exch vpt sub exch vpt sub vpt Square fill Bsquare } bind def
/S6 { BL [] 0 setdash 2 copy exch vpt sub exch vpt sub vpt vpt2 Rec fill Bsquare } bind def
/S7 { BL [] 0 setdash 2 copy exch vpt sub exch vpt sub vpt vpt2 Rec fill
       2 copy vpt Square fill
       Bsquare } bind def
/S8 { BL [] 0 setdash 2 copy vpt sub vpt Square fill Bsquare } bind def
/S9 { BL [] 0 setdash 2 copy vpt sub vpt vpt2 Rec fill Bsquare } bind def
/S10 { BL [] 0 setdash 2 copy vpt sub vpt Square fill 2 copy exch vpt sub exch vpt Square fill
       Bsquare } bind def
/S11 { BL [] 0 setdash 2 copy vpt sub vpt Square fill 2 copy exch vpt sub exch vpt2 vpt Rec fill
       Bsquare } bind def
/S12 { BL [] 0 setdash 2 copy exch vpt sub exch vpt sub vpt2 vpt Rec fill Bsquare } bind def
/S13 { BL [] 0 setdash 2 copy exch vpt sub exch vpt sub vpt2 vpt Rec fill
       2 copy vpt Square fill Bsquare } bind def
/S14 { BL [] 0 setdash 2 copy exch vpt sub exch vpt sub vpt2 vpt Rec fill
       2 copy exch vpt sub exch vpt Square fill Bsquare } bind def
/S15 { BL [] 0 setdash 2 copy Bsquare fill Bsquare } bind def
/D0 { gsave translate 45 rotate 0 0 S0 stroke grestore } bind def
/D1 { gsave translate 45 rotate 0 0 S1 stroke grestore } bind def
/D2 { gsave translate 45 rotate 0 0 S2 stroke grestore } bind def
/D3 { gsave translate 45 rotate 0 0 S3 stroke grestore } bind def
/D4 { gsave translate 45 rotate 0 0 S4 stroke grestore } bind def
/D5 { gsave translate 45 rotate 0 0 S5 stroke grestore } bind def
/D6 { gsave translate 45 rotate 0 0 S6 stroke grestore } bind def
/D7 { gsave translate 45 rotate 0 0 S7 stroke grestore } bind def
/D8 { gsave translate 45 rotate 0 0 S8 stroke grestore } bind def
/D9 { gsave translate 45 rotate 0 0 S9 stroke grestore } bind def
/D10 { gsave translate 45 rotate 0 0 S10 stroke grestore } bind def
/D11 { gsave translate 45 rotate 0 0 S11 stroke grestore } bind def
/D12 { gsave translate 45 rotate 0 0 S12 stroke grestore } bind def
/D13 { gsave translate 45 rotate 0 0 S13 stroke grestore } bind def
/D14 { gsave translate 45 rotate 0 0 S14 stroke grestore } bind def
/D15 { gsave translate 45 rotate 0 0 S15 stroke grestore } bind def
/DiaE { stroke [] 0 setdash vpt add M
  hpt neg vpt neg V hpt vpt neg V
  hpt vpt V hpt neg vpt V closepath stroke } def
/BoxE { stroke [] 0 setdash exch hpt sub exch vpt add M
  0 vpt2 neg V hpt2 0 V 0 vpt2 V
  hpt2 neg 0 V closepath stroke } def
/TriUE { stroke [] 0 setdash vpt 1.12 mul add M
  hpt neg vpt -1.62 mul V
  hpt 2 mul 0 V
  hpt neg vpt 1.62 mul V closepath stroke } def
/TriDE { stroke [] 0 setdash vpt 1.12 mul sub M
  hpt neg vpt 1.62 mul V
  hpt 2 mul 0 V
  hpt neg vpt -1.62 mul V closepath stroke } def
/PentE { stroke [] 0 setdash gsave
  translate 0 hpt M 4 {72 rotate 0 hpt L} repeat
  closepath stroke grestore } def
/CircE { stroke [] 0 setdash 
  hpt 0 360 arc stroke } def
/Opaque { gsave closepath 1 setgray fill grestore 0 setgray closepath } def
/DiaW { stroke [] 0 setdash vpt add M
  hpt neg vpt neg V hpt vpt neg V
  hpt vpt V hpt neg vpt V Opaque stroke } def
/BoxW { stroke [] 0 setdash exch hpt sub exch vpt add M
  0 vpt2 neg V hpt2 0 V 0 vpt2 V
  hpt2 neg 0 V Opaque stroke } def
/TriUW { stroke [] 0 setdash vpt 1.12 mul add M
  hpt neg vpt -1.62 mul V
  hpt 2 mul 0 V
  hpt neg vpt 1.62 mul V Opaque stroke } def
/TriDW { stroke [] 0 setdash vpt 1.12 mul sub M
  hpt neg vpt 1.62 mul V
  hpt 2 mul 0 V
  hpt neg vpt -1.62 mul V Opaque stroke } def
/PentW { stroke [] 0 setdash gsave
  translate 0 hpt M 4 {72 rotate 0 hpt L} repeat
  Opaque stroke grestore } def
/CircW { stroke [] 0 setdash 
  hpt 0 360 arc Opaque stroke } def
/BoxFill { gsave Rec 1 setgray fill grestore } def
end
}}%
\begin{picture}(3600,2160)(0,0)%
{\GNUPLOTspecial{"
gnudict begin
gsave
0 0 translate
0.100 0.100 scale
0 setgray
newpath
1.000 UL
LTb
150 300 M
0 63 V
0 1697 R
0 -63 V
810 300 M
0 63 V
0 1697 R
0 -63 V
1470 300 M
0 63 V
0 1697 R
0 -63 V
2130 300 M
0 63 V
0 1697 R
0 -63 V
2790 300 M
0 63 V
0 1697 R
0 -63 V
3450 300 M
0 63 V
0 1697 R
0 -63 V
1.000 UL
LTb
150 300 M
3300 0 V
0 1760 V
-3300 0 V
150 300 L
1.000 UL
LT0
150 2060 M
1 -1 V
2 0 V
1 -1 V
1 -1 V
2 -1 V
1 0 V
1 -1 V
2 -1 V
1 -1 V
1 0 V
2 -1 V
1 -1 V
1 -1 V
1 0 V
2 -1 V
1 -1 V
1 0 V
2 -1 V
1 -1 V
1 -1 V
2 0 V
1 -1 V
1 -1 V
2 -1 V
1 0 V
1 -1 V
2 -1 V
1 -1 V
1 0 V
2 -1 V
1 -1 V
1 -1 V
2 0 V
1 -1 V
1 -1 V
2 0 V
1 -1 V
1 -1 V
1 -1 V
2 0 V
1 -1 V
1 -1 V
2 -1 V
1 0 V
1 -1 V
2 -1 V
1 -1 V
1 0 V
2 -1 V
1 -1 V
1 0 V
2 -1 V
1 -1 V
1 -1 V
2 0 V
1 -1 V
1 -1 V
2 -1 V
1 0 V
1 -1 V
2 -1 V
1 -1 V
1 0 V
1 -1 V
2 -1 V
1 0 V
1 -1 V
2 -1 V
1 -1 V
1 0 V
2 -1 V
1 -1 V
1 -1 V
2 0 V
1 -1 V
1 -1 V
2 -1 V
1 0 V
1 -1 V
2 -1 V
1 -1 V
1 0 V
2 -1 V
1 -1 V
1 0 V
2 -1 V
1 -1 V
1 -1 V
1 0 V
2 -1 V
1 -1 V
1 -1 V
2 0 V
1 -1 V
1 -1 V
2 -1 V
1 0 V
1 -1 V
2 -1 V
1 0 V
1 -1 V
2 -1 V
1 -1 V
1 0 V
2 -1 V
1 -1 V
1 -1 V
2 0 V
1 -1 V
1 -1 V
2 -1 V
1 0 V
1 -1 V
1 -1 V
2 -1 V
1 0 V
1 -1 V
2 -1 V
1 0 V
1 -1 V
2 -1 V
1 -1 V
1 0 V
2 -1 V
1 -1 V
1 -1 V
2 0 V
1 -1 V
1 -1 V
2 -1 V
1 0 V
1 -1 V
2 -1 V
1 0 V
1 -1 V
2 -1 V
1 -1 V
1 0 V
1 -1 V
2 -1 V
1 -1 V
1 0 V
2 -1 V
1 -1 V
1 -1 V
2 0 V
1 -1 V
1 -1 V
2 -1 V
1 0 V
1 -1 V
2 -1 V
1 0 V
1 -1 V
2 -1 V
1 -1 V
1 0 V
2 -1 V
1 -1 V
1 -1 V
2 0 V
1 -1 V
1 -1 V
1 -1 V
2 0 V
1 -1 V
1 -1 V
2 0 V
1 -1 V
1 -1 V
2 -1 V
1 0 V
1 -1 V
2 -1 V
1 -1 V
1 0 V
2 -1 V
1 -1 V
1 -1 V
2 0 V
1 -1 V
1 -1 V
2 -1 V
1 0 V
1 -1 V
2 -1 V
1 0 V
1 -1 V
1 -1 V
2 -1 V
1 0 V
1 -1 V
2 -1 V
1 -1 V
1 0 V
2 -1 V
1 -1 V
1 -1 V
2 0 V
1 -1 V
1 -1 V
2 -1 V
1 0 V
1 -1 V
2 -1 V
1 0 V
1 -1 V
2 -1 V
1 -1 V
1 0 V
2 -1 V
1 -1 V
1 -1 V
1 0 V
2 -1 V
1 -1 V
1 -1 V
2 0 V
1 -1 V
1 -1 V
2 0 V
1 -1 V
1 -1 V
2 -1 V
1 0 V
1 -1 V
2 -1 V
1 -1 V
1 0 V
2 -1 V
1 -1 V
1 -1 V
2 0 V
1 -1 V
1 -1 V
2 -1 V
1 0 V
1 -1 V
1 -1 V
2 0 V
1 -1 V
1 -1 V
2 -1 V
1 0 V
1 -1 V
2 -1 V
1 -1 V
1 0 V
2 -1 V
1 -1 V
1 -1 V
2 0 V
1 -1 V
1 -1 V
2 -1 V
1 0 V
1 -1 V
2 -1 V
1 0 V
1 -1 V
2 -1 V
1 -1 V
1 0 V
1 -1 V
2 -1 V
1 -1 V
1 0 V
2 -1 V
1 -1 V
1 -1 V
2 0 V
1 -1 V
1 -1 V
2 -1 V
1 0 V
1 -1 V
2 -1 V
1 0 V
1 -1 V
2 -1 V
1 -1 V
1 0 V
2 -1 V
1 -1 V
1 -1 V
2 0 V
1 -1 V
1 -1 V
1 -1 V
2 0 V
1 -1 V
1 -1 V
2 -1 V
1 0 V
1 -1 V
2 -1 V
1 0 V
1 -1 V
2 -1 V
1 -1 V
1 0 V
2 -1 V
1 -1 V
1 -1 V
2 0 V
1 -1 V
1 -1 V
2 -1 V
1 0 V
1 -1 V
2 -1 V
1 -1 V
1 0 V
1 -1 V
2 -1 V
1 0 V
1 -1 V
2 -1 V
1 -1 V
1 0 V
2 -1 V
1 -1 V
1 -1 V
2 0 V
1 -1 V
1 -1 V
2 -1 V
1 0 V
1 -1 V
2 -1 V
1 -1 V
1 0 V
2 -1 V
1 -1 V
1 -1 V
2 0 V
1 -1 V
1 -1 V
1 0 V
2 -1 V
1 -1 V
1 -1 V
2 0 V
1 -1 V
1 -1 V
2 -1 V
1 0 V
1 -1 V
2 -1 V
1 -1 V
1 0 V
2 -1 V
1 -1 V
1 -1 V
2 0 V
1 -1 V
1 -1 V
2 0 V
1 -1 V
1 -1 V
2 -1 V
1 0 V
1 -1 V
1 -1 V
2 -1 V
1 0 V
1 -1 V
2 -1 V
1 -1 V
1 0 V
2 -1 V
1 -1 V
1 -1 V
2 0 V
1 -1 V
1 -1 V
2 -1 V
1 0 V
1 -1 V
2 -1 V
1 0 V
1 -1 V
2 -1 V
1 -1 V
1 0 V
2 -1 V
1 -1 V
1 -1 V
1 0 V
2 -1 V
1 -1 V
1 -1 V
2 0 V
1 -1 V
1 -1 V
2 -1 V
1 0 V
1 -1 V
2 -1 V
currentpoint stroke M
1 0 V
1 -1 V
2 -1 V
1 -1 V
1 0 V
2 -1 V
1 -1 V
1 -1 V
2 0 V
1 -1 V
1 -1 V
2 -1 V
1 0 V
1 -1 V
1 -1 V
2 -1 V
1 0 V
1 -1 V
2 -1 V
1 -1 V
1 0 V
2 -1 V
1 -1 V
1 0 V
2 -1 V
1 -1 V
1 -1 V
2 0 V
1 -1 V
1 -1 V
2 -1 V
1 0 V
1 -1 V
2 -1 V
1 -1 V
1 0 V
2 -1 V
1 -1 V
1 -1 V
1 0 V
2 -1 V
1 -1 V
1 -1 V
2 0 V
1 -1 V
1 -1 V
2 0 V
1 -1 V
1 -1 V
2 -1 V
1 0 V
1 -1 V
2 -1 V
1 -1 V
1 0 V
2 -1 V
1 -1 V
1 -1 V
2 0 V
1 -1 V
1 -1 V
2 -1 V
1 0 V
1 -1 V
1 -1 V
2 -1 V
1 0 V
1 -1 V
2 -1 V
1 0 V
1 -1 V
2 -1 V
1 -1 V
1 0 V
2 -1 V
1 -1 V
1 -1 V
2 0 V
1 -1 V
1 -1 V
2 -1 V
1 0 V
1 -1 V
2 -1 V
1 -1 V
1 0 V
2 -1 V
1 -1 V
1 -1 V
1 0 V
2 -1 V
1 -1 V
1 0 V
2 -1 V
1 -1 V
1 -1 V
2 0 V
1 -1 V
1 -1 V
2 -1 V
1 0 V
1 -1 V
2 -1 V
1 -1 V
1 0 V
2 -1 V
1 -1 V
1 -1 V
2 0 V
1 -1 V
1 -1 V
2 -1 V
1 0 V
1 -1 V
1 -1 V
2 -1 V
1 0 V
1 -1 V
2 -1 V
1 0 V
1 -1 V
2 -1 V
1 -1 V
1 0 V
2 -1 V
1 -1 V
1 -1 V
2 0 V
1 -1 V
1 -1 V
2 -1 V
1 0 V
1 -1 V
2 -1 V
1 -1 V
1 0 V
2 -1 V
1 -1 V
1 -1 V
1 0 V
2 -1 V
1 -1 V
1 -1 V
2 0 V
1 -1 V
1 -1 V
2 0 V
1 -1 V
1 -1 V
2 -1 V
1 0 V
1 -1 V
2 -1 V
1 -1 V
1 0 V
2 -1 V
1 -1 V
1 -1 V
2 0 V
1 -1 V
1 -1 V
2 -1 V
1 0 V
1 -1 V
1 -1 V
2 -1 V
1 0 V
1 -1 V
2 -1 V
1 -1 V
1 0 V
2 -1 V
1 -1 V
1 0 V
2 -1 V
1 -1 V
1 -1 V
2 0 V
1 -1 V
1 -1 V
2 -1 V
1 0 V
1 -1 V
2 -1 V
1 -1 V
1 0 V
2 -1 V
1 -1 V
1 -1 V
1 0 V
2 -1 V
1 -1 V
1 -1 V
2 0 V
1 -1 V
1 -1 V
2 -1 V
1 0 V
1 -1 V
2 -1 V
1 -1 V
1 0 V
2 -1 V
1 -1 V
1 0 V
2 -1 V
1 -1 V
1 -1 V
2 0 V
1 -1 V
1 -1 V
2 -1 V
1 0 V
1 -1 V
1 -1 V
2 -1 V
1 0 V
1 -1 V
2 -1 V
1 -1 V
1 0 V
2 -1 V
1 -1 V
1 -1 V
2 0 V
1 -1 V
1 -1 V
2 -1 V
1 0 V
1 -1 V
2 -1 V
1 -1 V
1 0 V
2 -1 V
1 -1 V
1 0 V
2 -1 V
1 -1 V
1 -1 V
1 0 V
2 -1 V
1 -1 V
1 -1 V
2 0 V
1 -1 V
1 -1 V
2 -1 V
1 0 V
1 -1 V
2 -1 V
1 -1 V
1 0 V
2 -1 V
1 -1 V
1 -1 V
2 0 V
1 -1 V
1 -1 V
2 -1 V
1 0 V
1 -1 V
2 -1 V
1 -1 V
1 0 V
1 -1 V
2 -1 V
1 -1 V
1 0 V
2 -1 V
1 -1 V
1 0 V
2 -1 V
1 -1 V
1 -1 V
2 0 V
1 -1 V
1 -1 V
2 -1 V
1 0 V
1 -1 V
2 -1 V
1 -1 V
1 0 V
2 -1 V
1 -1 V
1 -1 V
2 0 V
1 -1 V
1 -1 V
1 -1 V
2 0 V
1 -1 V
1 -1 V
2 -1 V
1 0 V
1 -1 V
2 -1 V
1 -1 V
1 0 V
2 -1 V
1 -1 V
1 -1 V
2 0 V
1 -1 V
1 -1 V
2 -1 V
1 0 V
1 -1 V
2 -1 V
1 -1 V
1 0 V
2 -1 V
1 -1 V
1 0 V
1 -1 V
2 -1 V
1 -1 V
1 0 V
2 -1 V
1 -1 V
1 -1 V
2 0 V
1 -1 V
1 -1 V
2 -1 V
1 0 V
1 -1 V
2 -1 V
1 -1 V
1 0 V
2 -1 V
1 -1 V
1 -1 V
2 0 V
1 -1 V
1 -1 V
2 -1 V
1 0 V
1 -1 V
1 -1 V
2 -1 V
1 0 V
1 -1 V
2 -1 V
1 -1 V
1 0 V
2 -1 V
1 -1 V
1 -1 V
2 0 V
1 -1 V
1 -1 V
2 -1 V
1 0 V
1 -1 V
2 -1 V
1 -1 V
1 0 V
2 -1 V
1 -1 V
1 0 V
2 -1 V
1 -1 V
1 -1 V
1 0 V
2 -1 V
1 -1 V
1 -1 V
2 0 V
1 -1 V
1 -1 V
2 -1 V
1 0 V
1 -1 V
2 -1 V
1 -1 V
1 0 V
2 -1 V
1 -1 V
1 -1 V
2 0 V
1 -1 V
1 -1 V
2 -1 V
1 0 V
1 -1 V
2 -1 V
1 -1 V
1 0 V
1 -1 V
2 -1 V
1 -1 V
1 0 V
2 -1 V
1 -1 V
1 -1 V
2 0 V
1 -1 V
1 -1 V
2 -1 V
currentpoint stroke M
1 0 V
1 -1 V
2 -1 V
1 -1 V
1 0 V
2 -1 V
1 -1 V
1 -1 V
2 0 V
1 -1 V
1 -1 V
2 -1 V
1 0 V
1 -1 V
1 -1 V
2 -1 V
1 0 V
1 -1 V
2 -1 V
1 -1 V
1 0 V
2 -1 V
1 -1 V
1 -1 V
2 0 V
1 -1 V
1 -1 V
2 -1 V
1 0 V
1 -1 V
2 -1 V
1 -1 V
1 0 V
2 -1 V
1 -1 V
1 -1 V
2 0 V
1 -1 V
1 -1 V
1 0 V
2 -1 V
1 -1 V
1 -1 V
2 0 V
1 -1 V
1 -1 V
2 -1 V
1 0 V
1 -1 V
2 -1 V
1 -1 V
1 0 V
2 -1 V
1 -1 V
1 -1 V
2 0 V
1 -1 V
1 -1 V
2 -1 V
1 0 V
1 -1 V
2 -1 V
1 -1 V
1 0 V
1 -1 V
2 -1 V
1 -1 V
1 0 V
2 -1 V
1 -1 V
1 -1 V
2 0 V
1 -1 V
1 -1 V
2 -1 V
1 0 V
1 -1 V
2 -1 V
1 -1 V
1 0 V
2 -1 V
1 -1 V
1 -1 V
2 0 V
1 -1 V
1 -1 V
2 -1 V
1 0 V
1 -1 V
1 -1 V
2 -1 V
1 0 V
1 -1 V
2 -1 V
1 -1 V
1 0 V
2 -1 V
1 -1 V
1 -1 V
2 0 V
1 -1 V
1 -1 V
2 -1 V
1 0 V
1 -1 V
2 -1 V
1 -1 V
1 0 V
2 -1 V
1 -1 V
1 -1 V
2 0 V
1 -1 V
1 -1 V
1 -1 V
2 0 V
1 -1 V
1 -1 V
2 -1 V
1 0 V
1 -1 V
2 -1 V
1 -1 V
1 0 V
2 -1 V
1 -1 V
1 -1 V
2 0 V
1 -1 V
1 -1 V
2 -1 V
1 0 V
1 -1 V
2 -1 V
1 -1 V
1 0 V
2 -1 V
1 -1 V
1 -1 V
1 0 V
2 -1 V
1 -1 V
1 -1 V
2 0 V
1 -1 V
1 -1 V
2 -1 V
1 0 V
1 -1 V
2 -1 V
1 -1 V
1 0 V
2 -1 V
1 -1 V
1 -1 V
2 -1 V
1 0 V
1 -1 V
2 -1 V
1 -1 V
1 0 V
2 -1 V
1 -1 V
1 -1 V
1 0 V
2 -1 V
1 -1 V
1 -1 V
2 0 V
1 -1 V
1 -1 V
2 -1 V
1 0 V
1 -1 V
2 -1 V
1 -1 V
1 0 V
2 -1 V
1 -1 V
1 -1 V
2 0 V
1 -1 V
1 -1 V
2 -1 V
1 0 V
1 -1 V
2 -1 V
1 -1 V
1 0 V
1 -1 V
2 -1 V
1 -1 V
1 0 V
2 -1 V
1 -1 V
1 -1 V
2 0 V
1 -1 V
1 -1 V
2 -1 V
1 0 V
1 -1 V
2 -1 V
1 -1 V
1 0 V
2 -1 V
1 -1 V
1 -1 V
2 0 V
1 -1 V
1 -1 V
2 -1 V
1 0 V
1 -1 V
1 -1 V
2 -1 V
1 0 V
1 -1 V
2 -1 V
1 -1 V
1 -1 V
2 0 V
1 -1 V
1 -1 V
2 -1 V
1 0 V
1 -1 V
2 -1 V
1 -1 V
1 0 V
2 -1 V
1 -1 V
1 -1 V
2 0 V
1 -1 V
1 -1 V
2 -1 V
1 0 V
1 -1 V
1 -1 V
2 -1 V
1 0 V
1 -1 V
2 -1 V
1 -1 V
1 0 V
2 -1 V
1 -1 V
1 -1 V
2 0 V
1 -1 V
1 -1 V
2 -1 V
1 0 V
1 -1 V
2 -1 V
1 -1 V
1 0 V
2 -1 V
1 -1 V
1 -1 V
2 -1 V
1 0 V
1 -1 V
1 -1 V
2 -1 V
1 0 V
1 -1 V
2 -1 V
1 -1 V
1 0 V
2 -1 V
1 -1 V
1 -1 V
2 0 V
1 -1 V
1 -1 V
2 -1 V
1 0 V
1 -1 V
2 -1 V
1 -1 V
1 0 V
2 -1 V
1 -1 V
1 -1 V
2 0 V
1 -1 V
1 -1 V
1 -1 V
2 -1 V
1 0 V
1 -1 V
2 -1 V
1 -1 V
1 0 V
2 -1 V
1 -1 V
1 -1 V
2 0 V
1 -1 V
1 -1 V
2 -1 V
1 0 V
1 -1 V
2 -1 V
1 -1 V
1 0 V
2 -1 V
1 -1 V
1 -1 V
2 0 V
1 -1 V
1 -1 V
1 -1 V
2 -1 V
1 0 V
1 -1 V
2 -1 V
1 -1 V
1 0 V
2 -1 V
1 -1 V
1 -1 V
2 0 V
1 -1 V
1 -1 V
2 -1 V
1 0 V
1 -1 V
2 -1 V
1 -1 V
1 0 V
2 -1 V
1 -1 V
1 -1 V
2 -1 V
1 0 V
1 -1 V
1 -1 V
2 -1 V
1 0 V
1 -1 V
2 -1 V
1 -1 V
1 0 V
2 -1 V
1 -1 V
1 -1 V
2 0 V
1 -1 V
1 -1 V
2 -1 V
1 0 V
1 -1 V
2 -1 V
1 -1 V
1 -1 V
2 0 V
1 -1 V
1 -1 V
2 -1 V
1 0 V
1 -1 V
1 -1 V
2 -1 V
1 0 V
1 -1 V
2 -1 V
1 -1 V
1 0 V
2 -1 V
1 -1 V
1 -1 V
2 -1 V
1 0 V
1 -1 V
2 -1 V
1 -1 V
1 0 V
2 -1 V
1 -1 V
1 -1 V
2 0 V
1 -1 V
1 -1 V
2 -1 V
1 0 V
1 -1 V
1 -1 V
2 -1 V
1 -1 V
1 0 V
2 -1 V
1 -1 V
1 -1 V
2 0 V
1 -1 V
1 -1 V
2 -1 V
currentpoint stroke M
1 0 V
1 -1 V
2 -1 V
1 -1 V
1 -1 V
2 0 V
1 -1 V
1 -1 V
2 -1 V
1 0 V
1 -1 V
2 -1 V
1 -1 V
1 0 V
1 -1 V
2 -1 V
1 -1 V
1 -1 V
2 0 V
1 -1 V
1 -1 V
2 -1 V
1 0 V
1 -1 V
2 -1 V
1 -1 V
1 0 V
2 -1 V
1 -1 V
1 -1 V
2 -1 V
1 0 V
1 -1 V
2 -1 V
1 -1 V
1 0 V
2 -1 V
1 -1 V
1 -1 V
1 0 V
2 -1 V
1 -1 V
1 -1 V
2 -1 V
1 0 V
1 -1 V
2 -1 V
1 -1 V
1 0 V
2 -1 V
1 -1 V
1 -1 V
2 -1 V
1 0 V
1 -1 V
2 -1 V
1 -1 V
1 0 V
2 -1 V
1 -1 V
1 -1 V
2 0 V
1 -1 V
1 -1 V
1 -1 V
2 -1 V
1 0 V
1 -1 V
2 -1 V
1 -1 V
1 0 V
2 -1 V
1 -1 V
1 -1 V
2 -1 V
1 0 V
1 -1 V
2 -1 V
1 -1 V
1 0 V
2 -1 V
1 -1 V
1 -1 V
2 -1 V
1 0 V
1 -1 V
2 -1 V
1 -1 V
1 0 V
1 -1 V
2 -1 V
1 -1 V
1 -1 V
2 0 V
1 -1 V
1 -1 V
2 -1 V
1 0 V
1 -1 V
2 -1 V
1 -1 V
1 -1 V
2 0 V
1 -1 V
1 -1 V
2 -1 V
1 0 V
1 -1 V
2 -1 V
1 -1 V
1 -1 V
2 0 V
1 -1 V
1 -1 V
1 -1 V
2 0 V
1 -1 V
1 -1 V
2 -1 V
1 -1 V
1 0 V
2 -1 V
1 -1 V
1 -1 V
2 0 V
1 -1 V
1 -1 V
2 -1 V
1 -1 V
1 0 V
2 -1 V
1 -1 V
1 -1 V
2 0 V
1 -1 V
1 -1 V
2 -1 V
1 -1 V
1 0 V
1 -1 V
2 -1 V
1 -1 V
1 -1 V
2 0 V
1 -1 V
1 -1 V
2 -1 V
1 0 V
1 -1 V
2 -1 V
1 -1 V
1 -1 V
2 0 V
1 -1 V
1 -1 V
2 -1 V
1 -1 V
1 0 V
2 -1 V
1 -1 V
1 -1 V
2 0 V
1 -1 V
1 -1 V
1 -1 V
2 -1 V
1 0 V
1 -1 V
2 -1 V
1 -1 V
1 -1 V
2 0 V
1 -1 V
1 -1 V
2 -1 V
1 0 V
1 -1 V
2 -1 V
1 -1 V
1 -1 V
2 0 V
1 -1 V
1 -1 V
2 -1 V
1 -1 V
1 0 V
2 -1 V
1 -1 V
1 -1 V
1 -1 V
2 0 V
1 -1 V
1 -1 V
2 -1 V
1 -1 V
1 0 V
2 -1 V
1 -1 V
1 -1 V
2 0 V
1 -1 V
1 -1 V
2 -1 V
1 -1 V
1 0 V
2 -1 V
1 -1 V
1 -1 V
2 -1 V
1 0 V
1 -1 V
2 -1 V
1 -1 V
1 -1 V
1 0 V
2 -1 V
1 -1 V
1 -1 V
2 -1 V
1 0 V
1 -1 V
2 -1 V
1 -1 V
1 -1 V
2 0 V
1 -1 V
1 -1 V
2 -1 V
1 -1 V
1 0 V
2 -1 V
1 -1 V
1 -1 V
2 -1 V
1 0 V
1 -1 V
2 -1 V
1 -1 V
1 -1 V
1 0 V
2 -1 V
1 -1 V
1 -1 V
2 -1 V
1 0 V
1 -1 V
2 -1 V
1 -1 V
1 -1 V
2 0 V
1 -1 V
1 -1 V
2 -1 V
1 -1 V
1 0 V
2 -1 V
1 -1 V
1 -1 V
2 -1 V
1 -1 V
1 0 V
2 -1 V
1 -1 V
1 -1 V
1 -1 V
2 0 V
1 -1 V
1 -1 V
2 -1 V
1 -1 V
1 0 V
2 -1 V
1 -1 V
1 -1 V
2 -1 V
1 0 V
1 -1 V
2 -1 V
1 -1 V
1 -1 V
2 -1 V
1 0 V
1 -1 V
2 -1 V
1 -1 V
1 -1 V
2 0 V
1 -1 V
1 -1 V
1 -1 V
2 -1 V
1 -1 V
1 0 V
2 -1 V
1 -1 V
1 -1 V
2 -1 V
1 0 V
1 -1 V
2 -1 V
1 -1 V
1 -1 V
2 -1 V
1 0 V
1 -1 V
2 -1 V
1 -1 V
1 -1 V
2 0 V
1 -1 V
1 -1 V
2 -1 V
1 -1 V
1 -1 V
1 0 V
2 -1 V
1 -1 V
1 -1 V
2 -1 V
1 -1 V
1 0 V
2 -1 V
1 -1 V
1 -1 V
2 -1 V
1 -1 V
1 0 V
2 -1 V
1 -1 V
1 -1 V
2 -1 V
1 -1 V
1 0 V
2 -1 V
1 -1 V
1 -1 V
2 -1 V
1 -1 V
1 0 V
1 -1 V
2 -1 V
1 -1 V
1 -1 V
2 -1 V
1 0 V
1 -1 V
2 -1 V
1 -1 V
1 -1 V
2 -1 V
1 0 V
1 -1 V
2 -1 V
1 -1 V
1 -1 V
2 -1 V
1 -1 V
1 0 V
2 -1 V
1 -1 V
1 -1 V
2 -1 V
1 -1 V
1 0 V
1 -1 V
2 -1 V
1 -1 V
1 -1 V
2 -1 V
1 -1 V
1 0 V
2 -1 V
1 -1 V
1 -1 V
2 -1 V
1 -1 V
1 -1 V
2 0 V
1 -1 V
1 -1 V
2 -1 V
1 -1 V
1 -1 V
2 -1 V
1 0 V
1 -1 V
2 -1 V
1 -1 V
1 -1 V
1 -1 V
2 -1 V
1 0 V
1 -1 V
2 -1 V
1 -1 V
1 -1 V
2 -1 V
1 -1 V
1 -1 V
2 0 V
currentpoint stroke M
1 -1 V
1 -1 V
2 -1 V
1 -1 V
1 -1 V
2 -1 V
1 -1 V
1 0 V
2 -1 V
1 -1 V
1 -1 V
2 -1 V
1 -1 V
1 -1 V
1 -1 V
2 -1 V
1 0 V
1 -1 V
2 -1 V
1 -1 V
1 -1 V
2 -1 V
1 -1 V
1 -1 V
2 -1 V
1 0 V
1 -1 V
2 -1 V
1 -1 V
1 -1 V
2 -1 V
1 -1 V
1 -1 V
2 -1 V
1 -1 V
1 -1 V
2 0 V
1 -1 V
1 -1 V
1 -1 V
2 -1 V
1 -1 V
1 -1 V
2 -1 V
1 -1 V
1 -1 V
2 -1 V
1 -1 V
1 0 V
2 -1 V
1 -1 V
1 -1 V
2 -1 V
1 -1 V
1 -1 V
2 -1 V
1 -1 V
1 -1 V
2 -1 V
1 -1 V
1 -1 V
2 -1 V
1 -1 V
1 -1 V
1 0 V
2 -1 V
1 -1 V
1 -1 V
2 -1 V
1 -1 V
1 -1 V
2 -1 V
1 -1 V
1 -1 V
2 -1 V
1 -1 V
1 -1 V
2 -1 V
1 -1 V
1 -1 V
2 -1 V
1 -1 V
1 -1 V
2 -1 V
1 -1 V
1 -1 V
2 -1 V
1 -1 V
1 -1 V
1 -1 V
2 -1 V
1 -1 V
1 -1 V
2 -1 V
1 -1 V
1 -1 V
2 -1 V
1 -1 V
1 -1 V
2 -1 V
1 -1 V
1 -2 V
2 -1 V
1 -1 V
1 -1 V
2 -1 V
1 -1 V
1 -1 V
2 -1 V
1 -1 V
1 -1 V
2 -1 V
1 -2 V
1 -1 V
1 -1 V
2 -1 V
1 -1 V
1 -1 V
2 -1 V
1 -2 V
1 -1 V
2 -1 V
1 -1 V
1 -1 V
2 -2 V
1 -1 V
1 -1 V
2 -1 V
1 -2 V
1 -1 V
2 -1 V
1 -1 V
1 -2 V
2 -1 V
1 -2 V
1 -1 V
2 -1 V
1 -2 V
1 -1 V
1 -2 V
2 -1 V
1 -2 V
1 -2 V
2 -2 V
1 -1 V
1 -2 V
2 -2 V
1 -3 V
1 -2 V
2 -4 V
1 -6 V
1 -1 V
2 -1 V
1 0 V
1 -1 V
2 -1 V
1 0 V
1 -1 V
2 -1 V
1 0 V
1 -1 V
2 -1 V
1 0 V
1 -1 V
1 -1 V
2 0 V
1 -1 V
1 -1 V
2 0 V
1 -1 V
1 -1 V
2 0 V
1 -1 V
1 -1 V
2 0 V
1 -1 V
1 -1 V
2 0 V
1 -1 V
1 -1 V
2 0 V
1 -1 V
1 -1 V
2 0 V
1 -1 V
1 -1 V
2 0 V
1 -1 V
1 -1 V
1 0 V
2 -1 V
1 -1 V
1 0 V
2 -1 V
1 -1 V
1 0 V
2 -1 V
1 -1 V
1 0 V
2 -1 V
1 -1 V
1 0 V
2 -1 V
1 -1 V
1 0 V
2 -1 V
1 -1 V
1 0 V
2 -1 V
1 -1 V
1 -1 V
2 0 V
1 -1 V
1 -1 V
1 0 V
2 -1 V
1 -1 V
1 0 V
2 -1 V
1 -1 V
1 0 V
2 -1 V
1 -1 V
1 0 V
2 -1 V
1 -1 V
1 0 V
2 -1 V
1 -1 V
1 -1 V
2 0 V
1 -1 V
1 -1 V
2 0 V
1 -1 V
1 -1 V
2 0 V
1 -1 V
1 -1 V
1 -1 V
2 0 V
1 -1 V
1 -1 V
2 0 V
1 -1 V
1 -1 V
2 0 V
1 -1 V
1 -1 V
2 -1 V
1 0 V
1 -1 V
2 -1 V
1 0 V
1 -1 V
2 -1 V
1 0 V
1 -1 V
2 -1 V
1 -1 V
1 0 V
2 -1 V
1 -1 V
1 0 V
1 -1 V
2 -1 V
1 -1 V
1 0 V
2 -1 V
1 -1 V
1 0 V
2 -1 V
1 -1 V
1 -1 V
2 0 V
1 -1 V
1 -1 V
2 -1 V
1 0 V
1 -1 V
2 -1 V
1 0 V
1 -1 V
2 -1 V
1 -1 V
1 0 V
2 -1 V
1 -1 V
1 -1 V
1 0 V
2 -1 V
1 -1 V
1 -1 V
2 0 V
1 -1 V
1 -1 V
2 -1 V
1 0 V
1 -1 V
2 -1 V
1 -1 V
1 0 V
2 -1 V
1 -1 V
1 -1 V
2 0 V
1 -1 V
1 -1 V
2 -1 V
1 -1 V
1 0 V
2 -1 V
1 -1 V
1 -1 V
1 0 V
2 -1 V
1 -1 V
1 -1 V
2 -1 V
1 0 V
1 -1 V
2 -1 V
1 -1 V
1 0 V
2 -1 V
1 -1 V
1 -1 V
2 -1 V
1 -1 V
1 0 V
2 -1 V
1 -1 V
1 -1 V
2 -1 V
1 0 V
1 -1 V
2 -1 V
1 -1 V
1 -1 V
1 -1 V
2 0 V
1 -1 V
1 -1 V
2 -1 V
1 -1 V
1 -1 V
2 -1 V
1 0 V
1 -1 V
2 -1 V
1 -1 V
1 -1 V
2 -1 V
1 -1 V
1 -1 V
2 -1 V
1 0 V
1 -1 V
2 -1 V
1 -1 V
1 -1 V
2 -1 V
1 -1 V
1 -1 V
1 -1 V
2 -1 V
1 -1 V
1 -1 V
2 -1 V
1 -1 V
1 -1 V
2 -1 V
1 -1 V
1 -1 V
2 -1 V
1 -1 V
1 -1 V
2 -1 V
1 -1 V
1 -2 V
2 -1 V
1 -1 V
1 -1 V
2 -1 V
1 -2 V
1 -1 V
2 -1 V
1 -1 V
1 -2 V
1 -1 V
2 -2 V
1 -1 V
1 -2 V
2 -1 V
1 -2 V
1 -2 V
2 -2 V
1 -2 V
1 -3 V
2 -3 V
currentpoint stroke M
1 -8 V
1 0 V
2 0 V
1 -1 V
1 0 V
2 -1 V
1 0 V
1 -1 V
2 0 V
1 -1 V
1 0 V
2 -1 V
1 0 V
1 0 V
1 -1 V
2 0 V
1 -1 V
1 0 V
2 -1 V
1 0 V
1 -1 V
2 0 V
1 -1 V
1 0 V
2 -1 V
1 0 V
1 0 V
2 -1 V
1 0 V
1 -1 V
2 0 V
1 -1 V
1 0 V
2 -1 V
1 0 V
1 -1 V
2 0 V
1 -1 V
1 0 V
1 0 V
2 -1 V
1 0 V
1 -1 V
2 0 V
1 -1 V
1 0 V
2 -1 V
1 0 V
1 -1 V
2 0 V
1 -1 V
1 0 V
2 -1 V
1 0 V
1 0 V
2 -1 V
1 0 V
1 -1 V
2 0 V
1 -1 V
1 0 V
2 -1 V
1 0 V
1 -1 V
1 0 V
2 -1 V
1 0 V
1 -1 V
2 0 V
1 -1 V
1 0 V
2 0 V
1 -1 V
1 0 V
2 -1 V
1 0 V
1 -1 V
2 0 V
1 -1 V
1 0 V
2 -1 V
1 0 V
1 -1 V
2 0 V
1 -1 V
1 0 V
2 -1 V
1 0 V
1 -1 V
1 0 V
2 -1 V
1 0 V
1 -1 V
2 0 V
1 0 V
1 -1 V
2 0 V
1 -1 V
1 0 V
2 -1 V
1 0 V
1 -1 V
2 0 V
1 -1 V
1 0 V
2 -1 V
1 0 V
1 -1 V
2 0 V
1 -1 V
1 0 V
2 -1 V
1 0 V
1 -1 V
1 0 V
2 -1 V
1 0 V
1 -1 V
2 0 V
1 -1 V
1 0 V
2 -1 V
1 0 V
1 -1 V
2 0 V
1 -1 V
1 0 V
2 -1 V
1 0 V
1 -1 V
2 0 V
1 -1 V
1 0 V
2 -1 V
1 0 V
1 -1 V
2 0 V
1 -1 V
1 0 V
1 -1 V
2 -1 V
1 0 V
1 -1 V
2 0 V
1 -1 V
1 0 V
2 -1 V
1 0 V
1 -1 V
2 0 V
1 -1 V
1 0 V
2 -1 V
1 0 V
1 -1 V
2 0 V
1 -1 V
1 -1 V
2 0 V
1 -1 V
1 0 V
2 -1 V
1 0 V
1 -1 V
1 0 V
2 -1 V
1 0 V
1 -1 V
2 -1 V
1 0 V
1 -1 V
2 0 V
1 -1 V
1 0 V
2 -1 V
1 -1 V
1 0 V
2 -1 V
1 0 V
1 -1 V
2 -1 V
1 0 V
1 -1 V
2 0 V
1 -1 V
1 -1 V
2 0 V
1 -1 V
1 0 V
1 -1 V
2 -1 V
1 0 V
1 -1 V
2 0 V
1 -1 V
1 -1 V
2 0 V
1 -1 V
1 -1 V
2 0 V
1 -1 V
1 -1 V
2 0 V
1 -1 V
1 -1 V
2 0 V
1 -1 V
1 -1 V
2 0 V
1 -1 V
1 -1 V
2 -1 V
1 0 V
1 -1 V
1 -1 V
2 -1 V
1 0 V
1 -1 V
2 -1 V
1 -1 V
1 -1 V
2 -1 V
1 0 V
1 -1 V
2 -1 V
1 -1 V
1 -1 V
2 -1 V
1 -1 V
1 -1 V
2 -1 V
1 -1 V
1 -1 V
2 -1 V
1 -1 V
1 -1 V
2 -2 V
1 -1 V
1 -1 V
1 -2 V
2 -1 V
1 -2 V
1 -2 V
2 -2 V
1 -2 V
1 -2 V
2 -3 V
1 -3 V
1 -4 V
2 -6 V
stroke
grestore
end
showpage
}}%
\put(1800,50){\makebox(0,0){$E_\beta -{\cal E}_0\ {\rm eV}$}}%
\put(3450,200){\makebox(0,0){0}}%
\put(2790,200){\makebox(0,0){-1}}%
\put(2130,200){\makebox(0,0){-2}}%
\put(1470,200){\makebox(0,0){-3}}%
\put(810,200){\makebox(0,0){-4}}%
\put(150,200){\makebox(0,0){-5}}%
\end{picture}%
\endgroup
 